\documentclass{llncs}
\usepackage{epstopdf}
\usepackage{wrapfig}
\usepackage{pbox}
\usepackage{llncsdoc}
\usepackage{mathtools}
\usepackage{enumerate}
\usepackage{lmodern}
\usepackage{algorithmic}
\usepackage{parskip}
\usepackage{amssymb}
\usepackage{layouts}
\usepackage{cite}
\usepackage{multicol}
\usepackage[T1]{fontenc}
\usepackage{graphicx}
\usepackage[tight]{subfigure}
\usepackage{breqn}
\usepackage[linesnumbered,ruled,vlined]{algorithm2e}

\newenvironment{keywords}{
       \list{}{\advance\topsep by0.35cm\relax\small
       \leftmargin=1cm
       \labelwidth=0.35cm
       \listparindent=0.35cm
       \itemindent\listparindent
       \rightmargin\leftmargin}\item[\hskip\labelsep
                                     \bfseries Keywords:]}
     {\endlist}

\begin{document}
\bibliographystyle{splncs03}

\newcounter{save}\setcounter{save}{\value{section}}
{\def\addtocontents#1#2{}%
\def\addcontentsline#1#2#3{}%
\def\markboth#1#2{}%
\title{Exemplar or Matching: Modeling DCJ Problems with Unequal Content Genome Data}

\author{Zhaoming Yin\inst{2}, Jijun Tang \inst{1,3*}, Stephen W. Schaeffer \inst{4} \and David A. Bader\inst{2}\thanks{Corresponding Authors}}

\institute{School of Computer Science and Technology, Tianjin University, China\and
School of Computational Science and Engineering,\\Georgia Institute of Technology, USA \and
Dept. of Computer Science and Engineering, University of South Carolina, USA \and
The Huck Institutes of Life Sciences, Pennsylvania State University, USA}

\maketitle
\begin{abstract}
The edit distance under the \emph{DCJ} model can be computed in linear time 
for genomes with equal content or with \emph{Indels}. But it becomes {\bf NP}-Hard in the presence of 
duplications, a problem largely unsolved especially when \emph{Indels} are considered.
In this paper, we compare two mainstream methods to deal with duplications and associate them with \emph{Indels}:
one by deletion, namely \emph{DCJ-Indel-Exemplar} distance;
versus the other by gene matching, namely \emph{DCJ-Indel-Matching} distance.
We design branch-and-bound algorithms with set of optimization methods to
compute exact distances for both. 
Furthermore, median problems are discussed in alignment with both of these distance methods, 
which are to find a median genome that minimizes distances between itself and three given genomes. 
Lin-Kernighan (\emph{LK}) heuristic is leveraged and powered up by sub-graph 
decomposition and search space reduction technologies to handle median computation.
A wide range of experiments are conducted on 
synthetic data sets and real data sets to show pros and cons of these two distance metrics per se, 
as well as putting them in the median computation scenario. 

\end{abstract}
\section{Introduction}
Over the last years, many distance metrics have been introduced to calculate 
the dissimilarity between two genomes by genome rearrangement \cite{Blin2004, bader1, BafnaP98, yanco05}. 
Among them, \emph{DCJ} distance is largely studied in recent years due to its capability to model various forms of rearrangement events, 
with a cheap cost of linear time computation.
However, when consiering duplications, the distance computation becomes {\bf NP}-hard \cite{ChaFerRiz06} and {\bf APX}-hard \cite{Angibaud09,ChenFZ12} 
for various distance models. There are two approaches to treat duplications, 
both are targeted at removing duplicated genes, so that existing linear algorithms can be utilized subsequently. 
The first approach identifies the so called exemplar genes\cite{Sankoff1999} in order to retain one copy gene in each duplicated gene family, 
while the other assigns one-to-one matching to every duplicated genes in each gene family \cite{shaoL12, shao13}.
Situated in the context of duplications, gene insertion and deletion (\emph{Indels}) 
are also important rearrangement events that results in unequal contents\cite{Brewer19991702}.
Pioneer works were conducted to study the sorting and distance computation by reversals with \emph{Indels} \cite{ El-Mabrouk2001}.
Later on, the \emph{DCJ-Indel} distance metric was introduced to take advantages of the \emph{DCJ} model. 
Braga et al \cite{Braga2010} proposed the first framework to compute the \emph{DCJ-Indel} distance; 
Compeau later simplified the problem with a much more elegant distance formula \cite{phillip1}. 
In this paper, we adapt the previous research results to design algorithms 
that procure the ability to handle both duplications and \emph{Indels} when computing \emph{DCJ} distance.

As evolutionary analysis generally involves more than two species, it is necessary to extend the above distances to deal with multiple genomes. 
Since three species form the smallest evoliutionary tree, it is critical to study the median problem, 
which is to construct a genome that minimizes the sum of distances from itself to the three input genomes\cite{BME1,Guillaume1}. 
The median problem is {\bf NP}-hard under most distance metrics \cite{udm1, Caprara1,weixu2, Bergeron1}.
Several exact algorithms have been implemented to solve the \emph{DCJ} median problems on both circular \cite{weixu1,weixu2}
and linear chromosomes \cite{weixu3,weixu4}.
Some heuristics are brought forth to improve the speed of median computation, such as linear programming (\emph{LP}) \cite{Caprara1},
local search \cite{Liris-3340},
evolutionary programming \cite{gao12},
or simply searching on one promising direction \cite{vaibhav1}.
All these algorithms are intended for solving median problems with equal content genomes, which are highly unrealistic in practice.
In this paper, we implement a Lin-Kernighan heuristic leveraging the 
aforementioned two distance metrics to compute \emph{DCJ} median when duplications and \emph{Indels} are considered.

\section{Background}
\subsection{Genome Rearrangement Events and their Graph Representations}
{\bf Genome Rearrangement Events}
The ordering of a genome can be changed through rearrangement events such as reversals and transpositions. 
Fig~\ref{rearrangement} shows examples of different events of a single chromosome (1 -2 3 4 -5 6 7).
In the examples, we use signed numbers to represent different genes and their orientations.
Genome rearrangement events involve with multiple combinatorial optimization problems and graph representation is common to abstract these problems.
In this part, we will address the foundations of using the breakpoint graph to abstract genome rearrangement events.

\begin{figure}[ht]
\begin{center}
\includegraphics[width=0.85\textwidth]{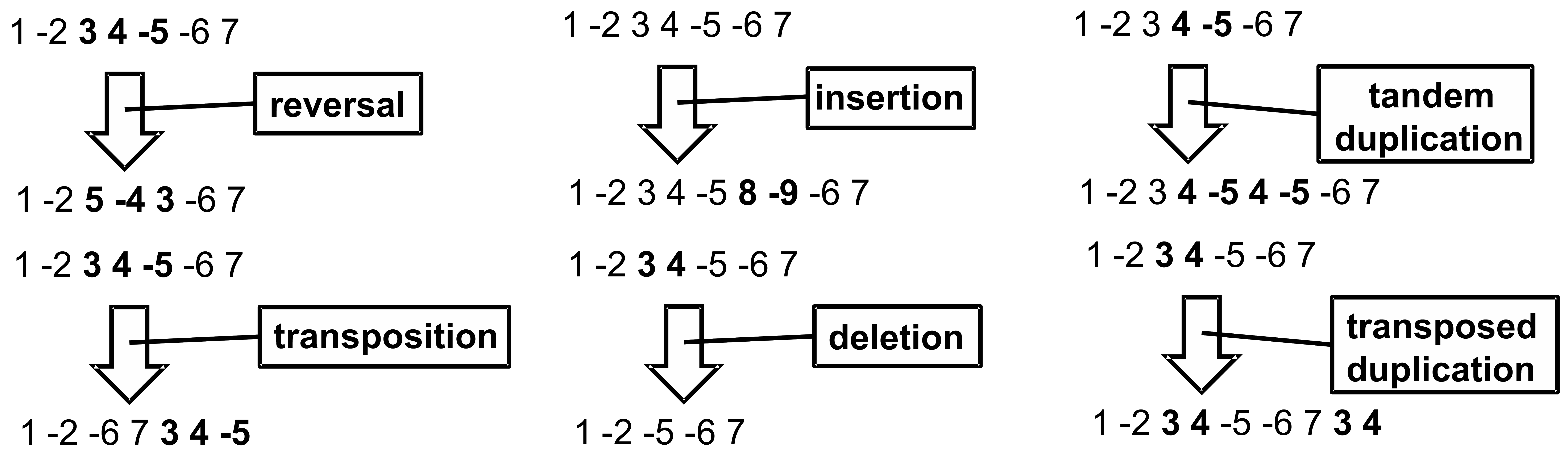}
    \caption{
	Example of different rearrangement events.
     }
\label{rearrangement}
\end{center}
\end{figure}

\begin{figure}[ht!]
    \begin{center}
	\subfigure[Example of \emph{BPG}]{
            \label{BPG}
            \includegraphics[width=0.70\textwidth]{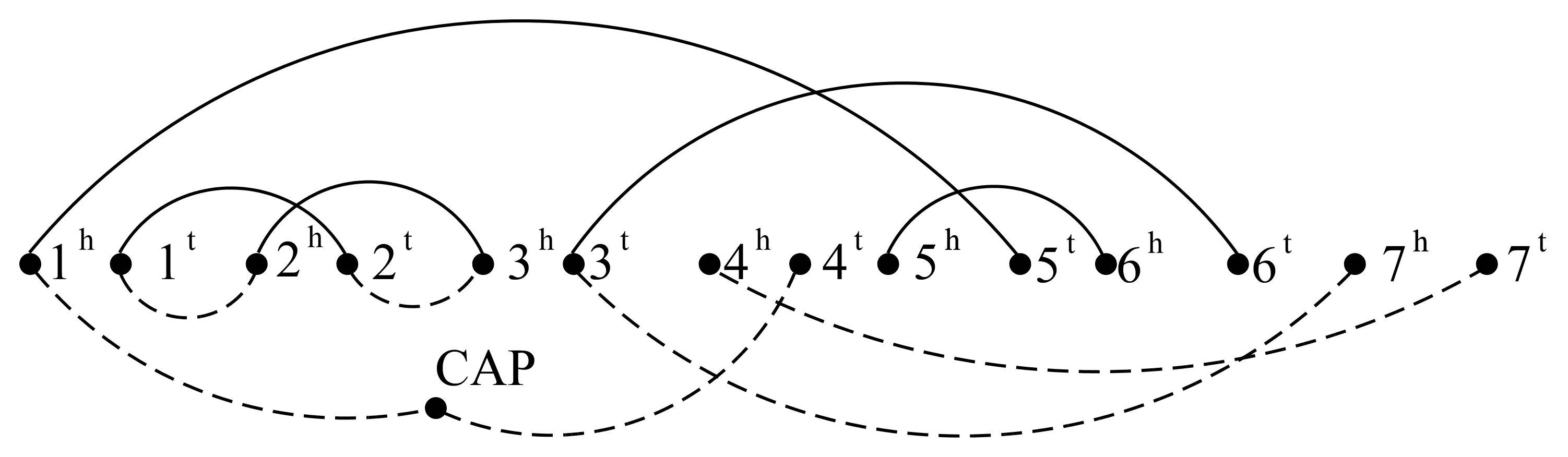}
        }%
        \subfigure[Example of \emph{DCJ}]{
           \label{dcj}
           \includegraphics[width=0.28\textwidth]{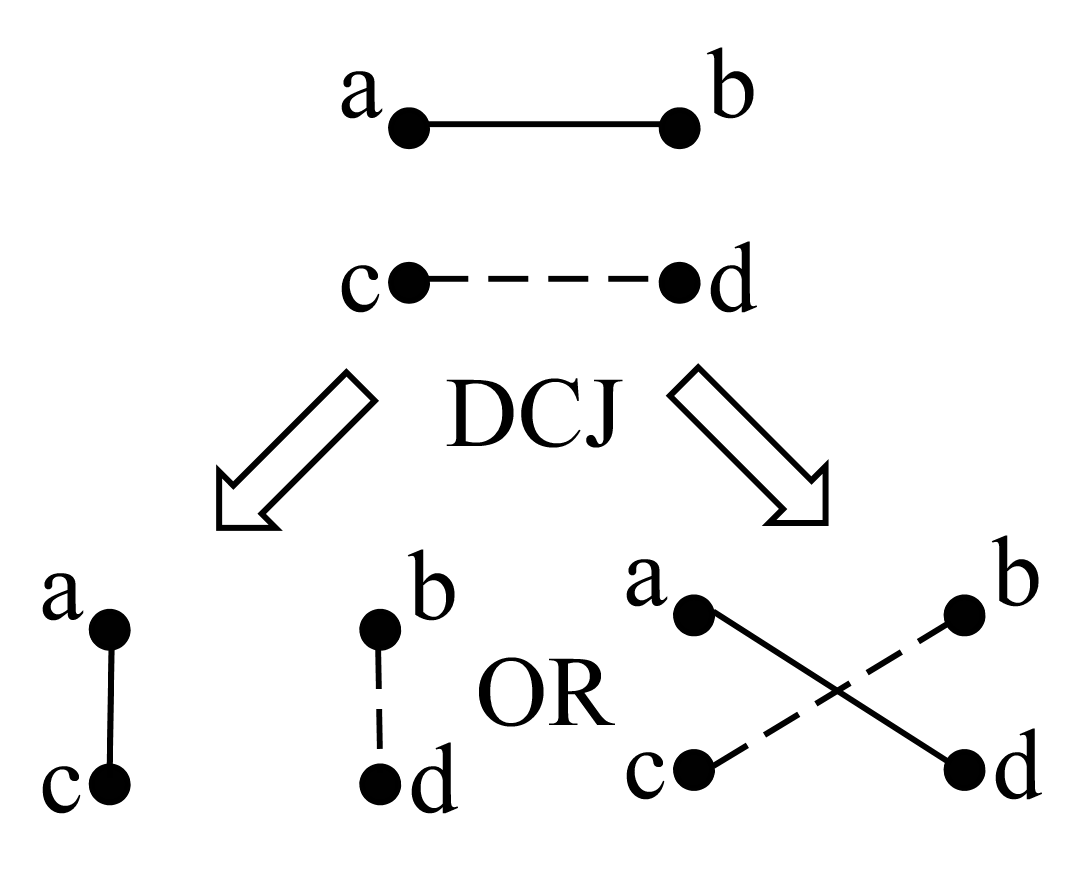}
        }
    \end{center}
    \caption{
	Examples of \emph{BPG}; and \emph{DCJ} operations.
     }
\label{distance}
\end{figure}

{\bf Breakpoint Graph}
Given an alphabet $\mathcal{A}$, two genomes $\Gamma$ and $\Pi$ are represented by two strings 
of signed ($+$ or $-$) numbers (representing genes) from $\mathcal{A}$.
Each gene $a\in\mathcal{A}$ is represented by a pair of vertices head $a_h$ and tail $a_t$; 
If $a$ is positive $a_h$ is putted in front of $a_t$, otherwise $a_t$ is putted in front of $a_h$.
For $a,b\in\mathcal{A}$, if $a,b\in\Gamma$ and are adjacent to each other,
their adjacent vertices will be connected by an edge.
For a telomere genes, if it exists in a circular chromosome, two end vertices will be connected by an edge; 
if it exists in a linear chromosome, two end vertices will be connected to a special vertex called \emph{CAP} vertex.
If we use one type of edges to represent adjacencies of gene order $\Gamma$ and another type of edges to represent adjacencies of gene order $\Pi$, 
the resulting graph with two types of edges is called breakpoint graph (\emph{BPG}).
Fig~\ref{BPG} shows the \emph{BPG} for gene order $\Gamma$ (1,-2,3,-6,5) (edge type: solid edges) 
which has one circular chromosome and $\Pi$ (1,2,3,7,4) (edge type: dashed edges) which has one linear chromosome.

{\bf \emph{DCJ} operation}
Double-cut and join (\emph{DCJ}) operations are able to simulate all rearrangement events.
In a \emph{BPG}, these operations cut two edges (within one genome) and rejoin them using two possible combinations of end vertices (shown in Fig~\ref{dcj}).

\subsection{Distance computation}
{\bf DCJ distance} \emph{DCJ} distance of genomes with the same content can be easily 
calculated by enumerating the number of cycles/paths in the \emph{BPG} \cite{yanco05},
which is of linear complexity.

{\bf DCJ-Indel distance} When \emph{Indels} are introduced in \emph{BPG}, with two genomes $\Gamma$ and $\Pi$, the vertices and edges of a closed walk form a cycle.  
In Fig~\ref{BPG}, the walk $(1^t, (1^t;2^h), 2^h,$ $(2^h;3^h), 3^h, (3^h;2^t), 2^t,$ $(2^t;1^t) , 1^t)$ is a cycle.
A vertex $v$ is $\pi$-$open$ $(\gamma$-$open)$ if $v \not\in \Gamma$ ($v \not\in \Pi$). 
An unclosed walk in \emph{BPG} is a path. Based on different kinds of ends points of paths, 
we can classify paths into different types. If the two ends of a path are \emph{CAP} vertices, 
we simply denote this path as $p^0$. If a path is ended by one open vertex and one \emph{CAP}, we
denote it as $p^\pi$ $(p^\gamma)$. If a path is ended by two open vertices, 
we denote it by the types of its two open vertices: for instance, 
$p^{\pi,\gamma}$ represents a path that ends with a $\pi$-$open$ vertex and a $\gamma$-$open$ vertex. In Fig~\ref{BPG}, 
the walk $(5^t,(5^t;1^h),1^h,(1^h;CAP),CAP)$ is a $p^\gamma$ path 
and the walk $(6^t,(6^t;3^t),3^t,(3^t;7^h),7^h)$ is a $p^{\gamma,\pi}$ path. 
A path is even (odd), if it contains even (odd) number of edges. 
In \cite{phillip1}, if  $|\mathcal{A}|=N$ the \emph{DCJ} distance between two genomes with \emph{Indels} but without duplications 
is calculated by equation (\ref{indeldis}). 
We call this distance \emph{DCJ-Indel} distance.
From this equation, we can easily get the \emph{DCJ-Indel} distance between $\Gamma$ and $\Pi$ in Fig~\ref{BPG} as $4$.
\begin{dmath}
d_{indel}(\Gamma,\Pi) = N - [|c|+|p^{\pi,\pi}|+|p^{\gamma,\gamma}|+\lfloor p^{\pi,\gamma}\rfloor] \\
 +\frac{1}{2}(|p_{even}^0| + min(|p_{odd}^{\pi}|,|p_{even}^{\pi}|)+min(|p_{odd}^{\gamma}|,|p_{even}^{\gamma}|) + \delta)
\label{indeldis}
\end{dmath}

Where $\delta=1$ only if $p^{\pi,\gamma}$ is odd and either 
$|p_{odd}^{\pi}| > |p_{even}^{\gamma}|,|p_{odd}^{\gamma}|>|p_{even}^{\gamma}|$ 
or  $|p_{odd}^{\pi}| < |p_{even}^{\gamma}|,|p_{odd}^{\gamma}|<|p_{even}^{\gamma}|$; 
Otherwise, $\delta=0$.

{\bf DCJ-Exemplar(Matching) distance}
There are in general two approaches to cope with duplicated genes. 
One is by removing all but keeping one copy in a gene family to generate an exemplar pair\cite{Sankoff1999}  
and the other is by relabeling duplicated genes to ensure that every duplicated gene has unique number\cite{shao13, shaoL12}. 
Both of these two distances can be computed with \emph{BPG} using branch-and-bound methods.
For both of the distance metrics, the upper bound can be easily derived by assigning an arbitrary mapping to two genomes then computing their mutual distance.
In paper \cite{Sankoff1999} regarding exemplar distance, it's proved that by removing all occurrences of unfixed duplicated gene families, 
the resulting distance is monotony decreasing, hence the resulting distance can be served as a lower bound.
In paper \cite{ChenZFNZLJ05} regarding matching distance, the authors proposed a way for computing lower bounds by measuring the number of breakpoints between two genomes, 
which might not directly imply the lower bound between genomes with \emph{Indels}. 
However, it is still possible to use this method to find a `relaxed' lower bound.

{\bf Distance Estimation} Note that mathematically optimized distance might not reflect the true number of biological events, 
thus several estimation methods such as \emph{EDE} or \emph{IEBP} are used to rescale 
these computed distances \cite{MoretWWW01} to better fit true evolutionary history.

\subsection{Median Computation}
\begin{figure}[ht!]
    \begin{center}
        \subfigure[\emph{MBG}]{
            \label{mbg}
            \includegraphics[width=0.25\textwidth]{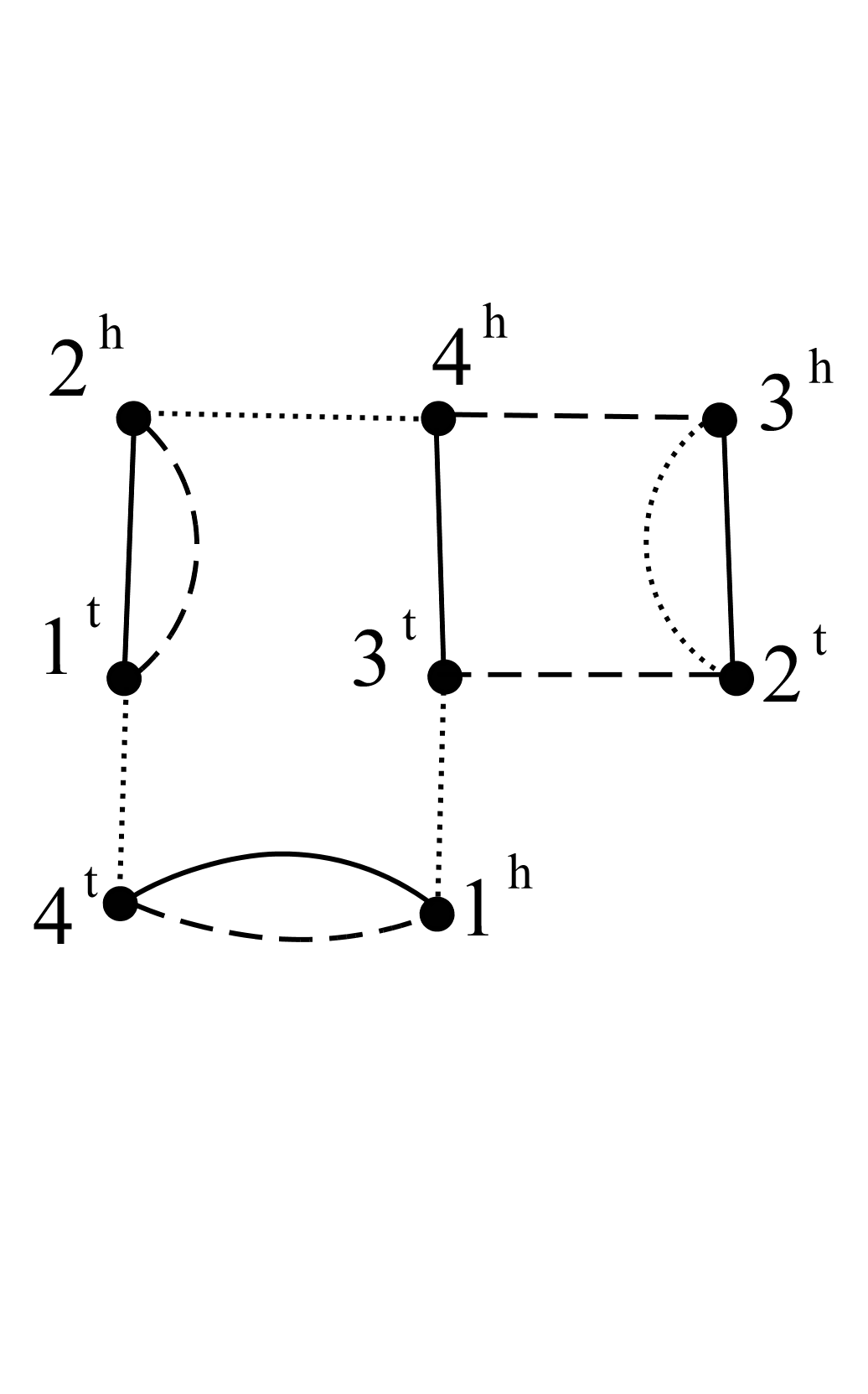}
        }%
        \subfigure[\emph{0-matching}]{
           \label{zmat}
           \includegraphics[width=0.25\textwidth]{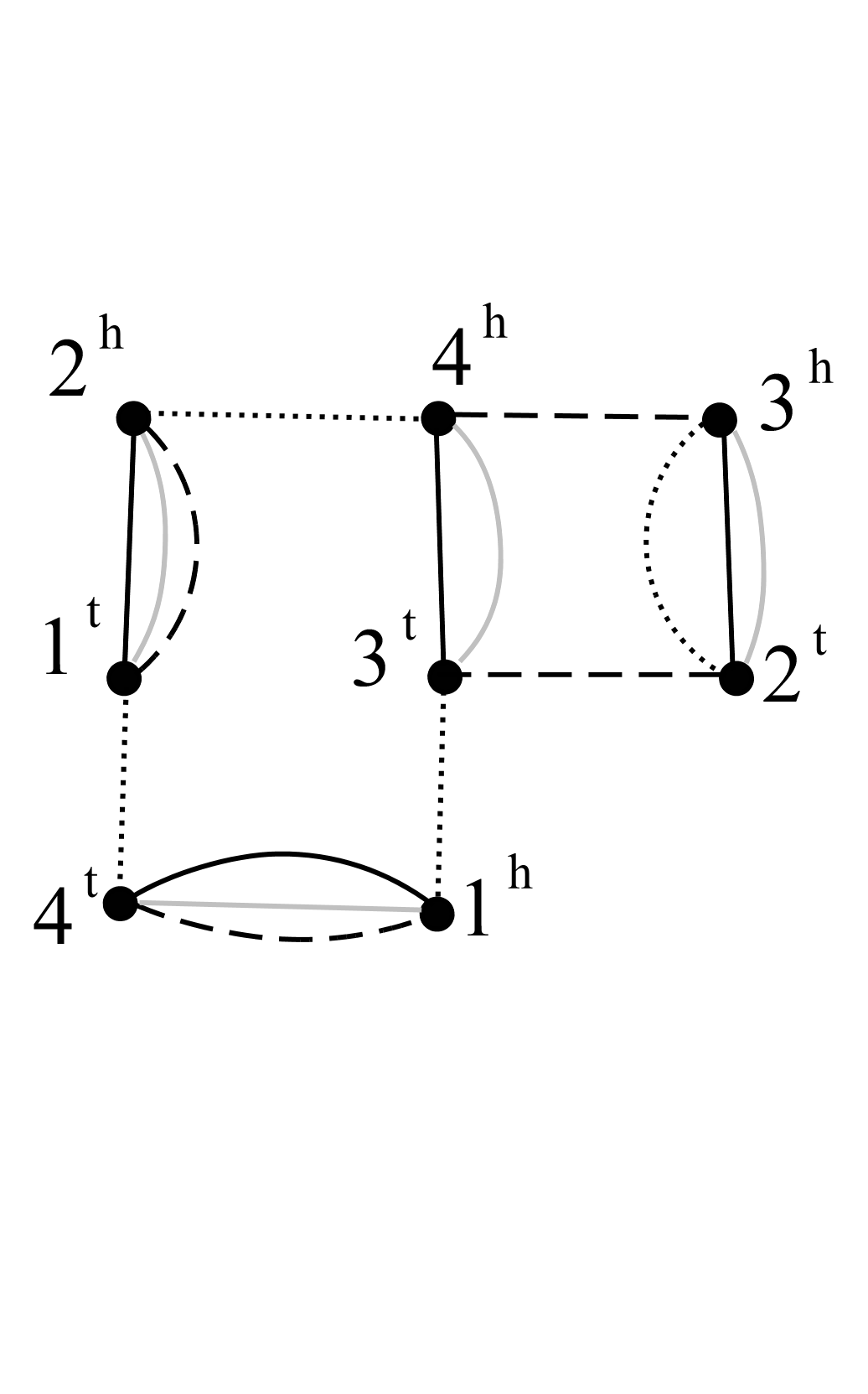}
        }
        \subfigure[Adequate subgraph and edge shrinking]{
           \label{as}
           \includegraphics[width=0.25\textwidth]{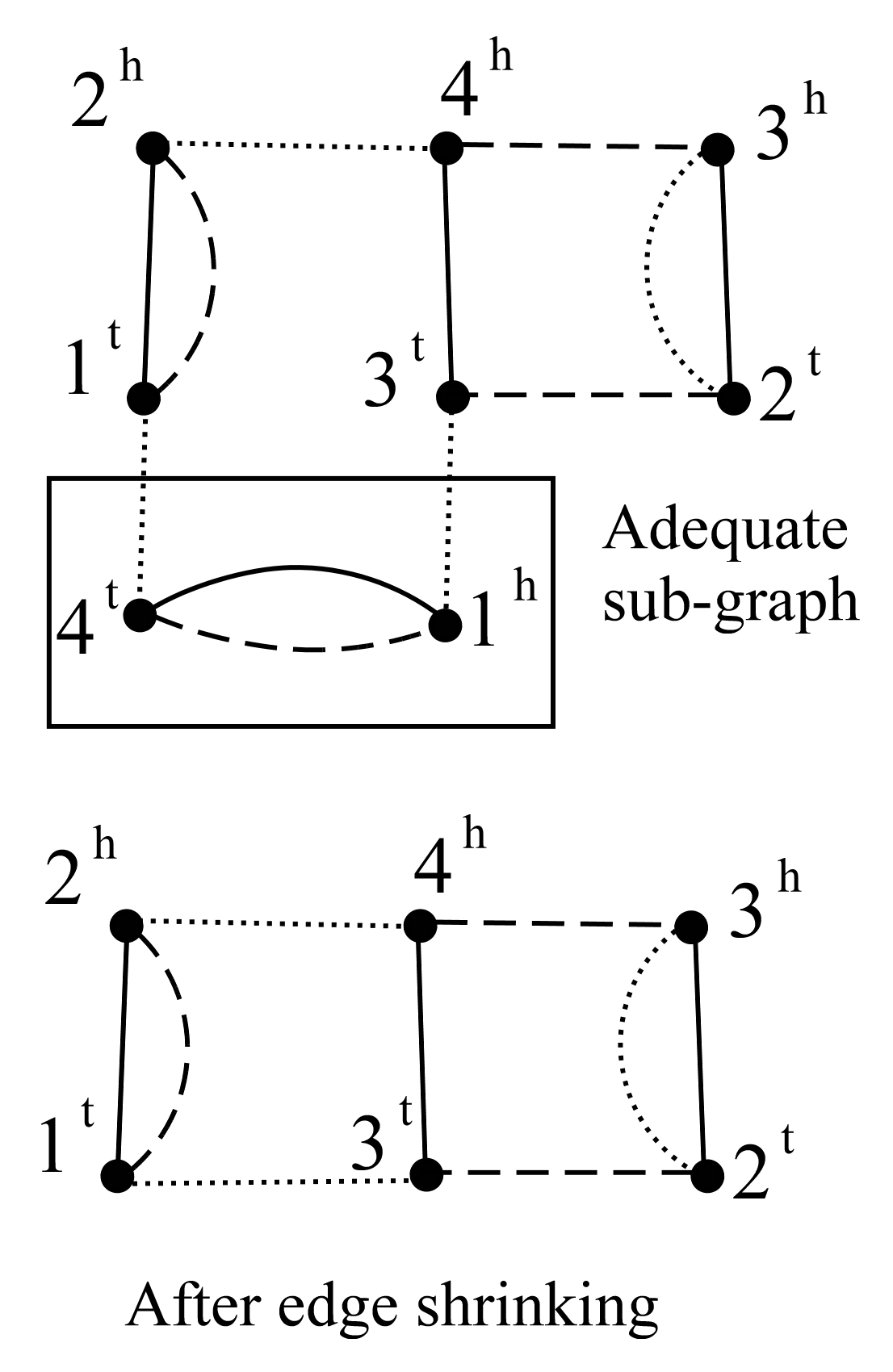}
        }
    \end{center}
    \caption{
       (top) Examples of \emph{MBG} with three input genomes: (1,2,3,4) (solid edges); (1,2,-3,4) (dashed edges) and (2,3,1,-4) (dotted edges).; 
			(middle) 0-matching operation; (bottom) edge shrinking operations. 
     }
\label{distance}
\end{figure}

If there are three given genomes, the graph constructed by pre-defined \emph{BPG} rule 
is called a Multiple Breakpoint Graph (\emph{MBG}). 
Figure~\ref{mbg} shows an example of \emph{MBG} with three input genomes.
When there are only equal content genomes, the \emph{DCJ} median problem can be briefly described by 
finding a maximum matching (which is called  $0$-$matching$) in \emph{MBG}. 
Figure~\ref{zmat} shows an example of $0$-$matching$ which is represented by gray edges.
In \cite{weixu1}, it is proven that a type of sub-graph called adequate sub-graph (\emph{AS}) 
could be used to decompose the graph with edge shrinking operations, which are shown in Figure~\ref{as}.
Unfortunately, there is no branch-and-bound based median algorithm that deals with unequal content genomes. 
In the following section, we will show that it is actually difficult to design such algorithm.

\begin{figure}[ht]
\begin{center}
\includegraphics[width=0.70\textwidth]{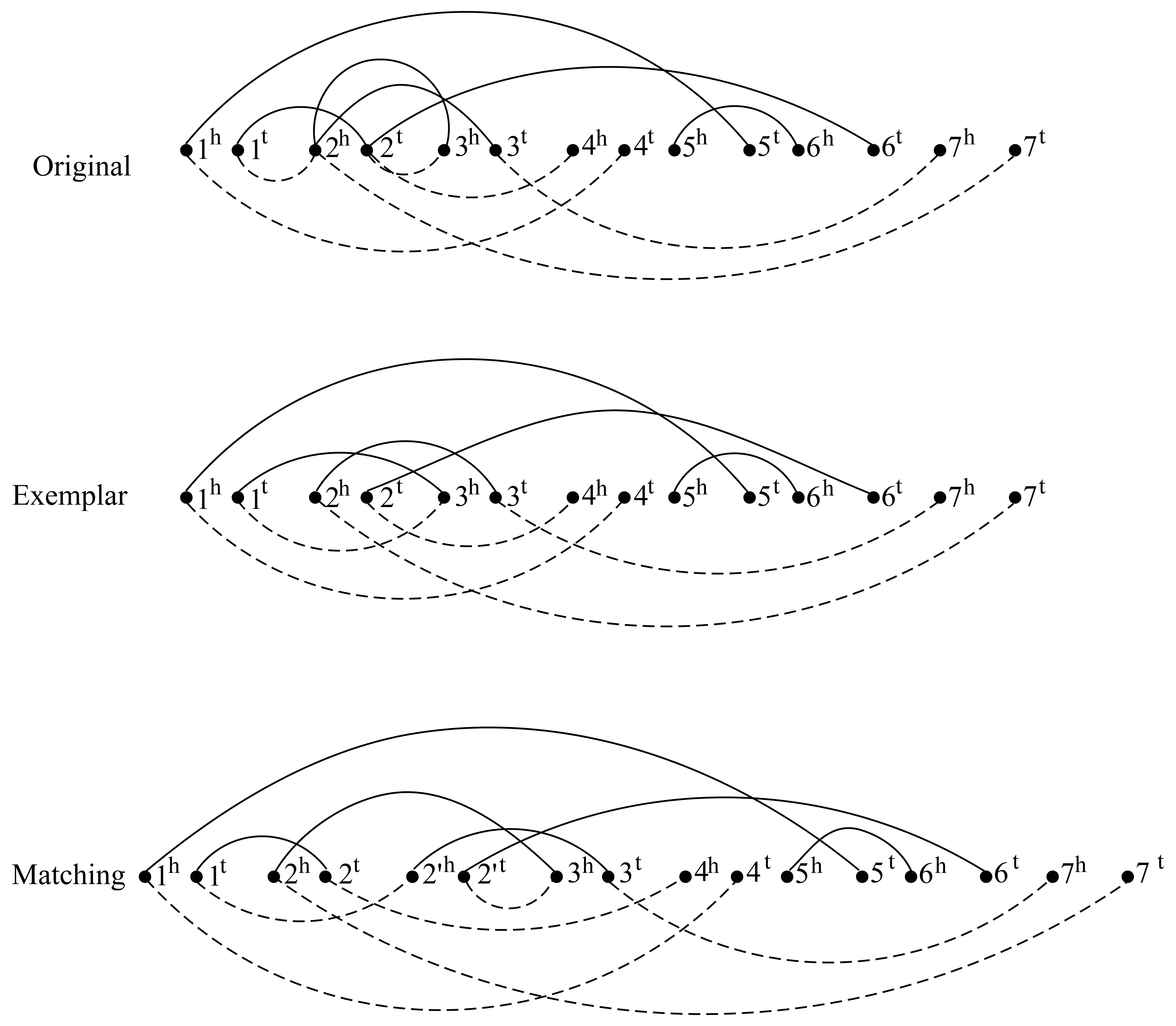}
\caption{
    Examples of exemplar and matching distance in the form of \emph{BPG} representation.
}
\label{exem_matching}
\end{center}
\end{figure}

\section{Approaches}
\subsection{Proposed Distance Metrics}
We have discussed \emph{DCJ}, \emph{DCJ-Indel} and \emph{DCJ-Exemplar(Matching)} distances, here we formally define the \emph{DCJ-Indel-Exemplar(Matching)} distances as follows:

{\bf Definition 1.} An \emph{exemplar} string is constructed by deleting all but one occurrence of each gene family. 
Among all possible exemplar strings, the minimum distance that one exemplar string returns is the \emph{DCJ-Indel-Exemplar} distance.

{\bf Definition 2.} A \emph{matching} string is constructed by assigning a one-to-one mapping to each occurrence of 
genes in a gene family and relabel them to distinct markers.
Among all possible matching strings, the minimum distance that one matching string returns is the \emph{DCJ-Indel-Matching} distance.

Figure~\ref{exem_matching} shows examples of \emph{BPG} representation of exemplar mapping from genome $\Gamma$ 
(1, -2, 3, 2, -6, 5) and genome $\Pi$ (1, 2, 3, 7, 2, 4) to $\Gamma$ (1, 3, 2, -6, 5) 
and genome $\Pi$ (1, 3, 7, 2, 4), and a matching that mapping from genome $\Gamma$ 
(1, -2, 3, 2, -6, 5) and genome $\Pi$ (1, 2, 3, 7, 2, 4) to 
$\Gamma$ (1, -2, 3, 2', -6, 5) and genome $\Pi$ (1, 2', 3, 7, 2, 4).

We can use branch-and-bound methods which are applied in \emph{DCJ-Exemplar} \emph{(Matching)} distances to solve these two distances.

\subsection{Optimization Methods}
{\bf Optimal Assignments} Although branch-and-bound algorithms are based on enumerating the number of cycles/path in \emph{BPG}, 
it is not necessary to enumerate every component in the graph, 
as both \cite{shao13, ChenZFNZLJ05} indicated that there are some specific patterns in \emph{BPG} which can be fixed before the distance computation.
In this paper, we will extend their result in our optimization methods for \emph{DCJ-Indel-Exemplar(Matching)} distances.

To begin with, we define some terms for future explanation.
There are two categories of vertices in a \emph{BPG}: one connects exactly one edge of each edge type 
(in this paper edge types are expressed by such as dotted, dashed edges etc.), they are called \emph{regular} vertices;
the other connects fewer or more than one edges of each edge type, they are called \emph{irregular} vertices.
A subgraph in a \emph{BPG} that only contains regular vertices is defined as \emph{regular subgraph}, while 
one that contains irregular vertices is defined as \emph{irregular subgraph}.
In \emph{BPG} with two genomes $\Gamma$ and $\Pi$, vertices and edges of a closed walk form a cycle.

\begin{theorem}
In a \emph{BPG}, an irregular subgraph which is a cycle of length 2 can be fixed before computation without losing accuracy. 
\end{theorem}

\begin{proof}
Without loss of generality, the proof is sound for both \emph{DCJ-Indel-Exemplar} and \emph{DCJ-Indel-Matching} distances. 
We prove the theorem under two cases: 
\begin{enumerate}
\item for the subgraph in the component which only contains cycles, 
this is a case that is exactly the same as mentioned in \cite{shao13}, proof.
\item for the subgraph in the component which contains paths, since no type of the paths has count more than one 
(which is the count of a cycle), following the similar proof strategy in \cite{shao13}, we can get the same conclusion. $\Box$ 
\end{enumerate} 

\end{proof}

{\bf Adopting Morph Graph Methods to Condense \emph{BPG}} 
If a gene family has multiple copies of the gene, its corresponding two vertices ($head$ and $tail$) in the \emph{BPG} will have degree of more than one.
In contrary, vertex representations of those singleton genes always have degree of one or zero.
Once an `exemplar' or `matching' is fixed, only edges incident to vertices that have degree of more than one have been changed.
We can view the computation of exemplar or matching distance as the process of morphing (or streaming) 
\cite{YinTSB13} the \emph{BPG} in order to find an ad hoc shape of the \emph{BPG} that achieves optimality.
Following this hint, we can bridge out all vertices that are stable and just investigate these dynamically changing vertices without lossing accuracy. 
Suppose there are $V$ vertices in the \emph{BPG}, where $V_s$ are stable and $V_d$ are dynamic, the asymptotic speedup for this morph \emph{BPG} strategy 
will be $O(\frac{V}{V_d})$.

{\bf Harness the Power of Divide-and-Conquer Approach to Reduce the Problem Space}
In the paper by Nguyen et al \cite{Nguyen2005}, the authors proposed a divide and conquer method to quickly calculate the exemplar distance.
Inspired by their idea, we propose the following divide-and-conquer method to compute the above two distances based on the \emph{BPG}.
We have the follow observation:

\begin{theorem}
The \emph{DCJ-Indel-Exemplar (Matching)} distance is optimal \emph{iff} 
the choices of exemplar edges (cycle decomposition) in each connected components of \emph{BPG} are optimal. 
\end{theorem}

\begin{proof}
Since it's obvious that for regular connected component of \emph{BPG}, there is only one choice of edges, the proof under this case is trivial.
For irregular connected component of \emph{BPG}, we prove by contrary: suppose there is another edge selection that can result in a better distance,
based on the corresponding \emph{BPG}, there must be at least one connected component that has a better edge selection, 
replacing it with a better edge selection will result in a better distance, which violates the assumption. $\Box$ 
\end{proof}

\begin{algorithm}[H]
\KwIn{$G_1$ and $G_2$ }
\KwOut{Minimum distance $d$}
optimization methods on $G_1$, $G_2$\;
$G_1^',G_2^' \gets $randomly init exemplar(matching) of all duplicated genes of $G_1$, $G_2$\;
$G_1^*,G_2^* \gets $remove all duplicated genes of $G_1$, $G_2$\;
$min\_ub \gets DCJIndel(G_1^',G_2^')$ \;
$min\_lb \gets DCJIndel(G_1^*,G_2^*)$ \;
Init search list $L$ of size $min\_ub - min\_lb$ and insert $G_1,G_2$\;
\While{$min\_ub > min\_lb$}{
    $G_1^{+},G_2^{+} \gets$ pop from $L[min\_lb]$\;
    \For{$pair \in$ all mappings of next available duplicated gene} {
        $G_1^{+},G_2^{+} \gets$ $G_1^{+},G_2^{+}$ fix the exemplar(matching) of $pair$ \;
        $G_1^{+'},G_2^{+'} \gets $randomly init exemplar(matching) of rest duplicated genes $G_1^{+}$, $G_2^{+}$\;
        $G_1^{+*},G_2^{+*} \gets $remove rest duplicated genes $G_1^+$, $G_2^+$\;
        $ub \gets DCJIndel(G_1^{+'},G_2^{+'})$ \;
        $lb \gets DCJIndel(G_1^{+*},G_2^{+*})$ \;
        \uIf{$lb > min\_ub$} {
            discard $G_1^+,G_2^+$
        }
        \uIf{$ub < min\_ub$} {
            $min\_ub = ub$\;
        }
        \uElseIf{$ub = max\_lb$} {
            \Return {$d=ub$} \;
        }
        \uElse{
            insert $G_1^{+},G_2^{+}$ into $L[lb]$
        }
    }
}
\Return{$d=min\_lb$}\;
\caption{{\sc DCJIndelExem(Matc)Distance}}
\label{algo:dist}
\end{algorithm}

Combining three optimization methods in tandem with the branch-and-bound framework, 
we can summarize our algorithm to compute \emph{DCJ-Indel-Exemplar} \emph{(Matching)} distance as outlined in Algorithm~\ref{algo:dist}.

\subsection{Adapting \emph{Lin-Kernighan} Heuristic to Find the Median Genome}
{\bf Problem Statement}
Not surprisingly, finding the median genome that minimizes the \emph{DCJ-Indel-Exemplar(Matching)} distance is challenging. 
To begin with, given three input genomes, there are multiple choices of possible gene content selections for the median; 
however, since identifying gene content is simpler and there exists very accurate and fast methods to fulfil the task \cite{feihu3}, 
we are more interested on a relaxed version of the median problem that assumes known gene content on the median genome. 
Which is formally defined as:

{\bf Definition:} Given the gene content of a median genome, and gene orders of three input genomes.
Find an adjacency of the genes of the median genome that minimize the \emph{DCJ-Indel-Exemplar(Matching)} distance between the median genome and the three input genomes.

The \emph{DCJ-Indel-Exemplar(Matching)} median problem is not even in the class of {\bf NP} 
because there is no polynomial time algorithm to verify the results. 
It is hard to design an exact branch-and-bound algorithm for the \emph{DCJ-Indel-Exemplar(Matching)} 
median problem mainly because the \emph{DCJ-Indel} distance violates the property of 
triangular inqueality which is required for a distance metrics \cite{YancopoulosF08}. 
Furthermore, when there are duplicated genes in a genome,  
it is possible that there are multiple edges of the same type connecting to the same vertex of a \emph{0-matching}, 
which leads to ambiguity in the edge shrinking step and makes the followed branch-and-bound search process very complicated and extremely hard to implement.
To overcome these problems, we provide an adaption of Lin-Kernighan (\emph{LK}) heuristic to help solving this challenging problem.

{\bf Design of the \emph{Lin-Kernighan} Heuristic}
The \emph{LK} heuristic can generally be divided into two steps: initialize the 
$0$-$matching$ for the median genome, and \emph{LK} search to get the result.

The initialization problem can be described as: given the gene contents of three input genomes, 
find the gene content of the median genome that minimizes the sum of the number of \emph{Indels} 
and duplications operations required to transfer the median gene content to the gene contents of the other three genomes.
In this paper, we design a very simple rule to initialize the median gene content: 
given the counts of each gene family occurred in the three genomes, 
if two or three counts are the same, we simply select this count as the number of occurrence of 
the gene family in the median genome; if all three counts are different, 
we select the median count as the number of occurrence of the gene family in the median genome. 

After fixing the gene content for the median genome, we randomly set up the \emph{0-matching} in the \emph{MBG}. 
The followed \emph{LK} heuristic selects two \emph{0-matching} edges on the \emph{MBG} of a given search node 
and performs a \emph{DCJ} operation, obtaining the \emph{MBG} of a neighboring search node. 
We expand the search frontier by keeping all neighboring search nodes to up until the search level $L1$. 
Then we only examine and add the most promising neighbors to the search list until level $L2$. 
The search is continued when there is a neighbor solution yielding a better median score. 
This solution is then accepted and a new search is initialized from the scratch.
The search will be terminated if there are no improvement on the result as the search level
limits have been reached and all possible neighbors have been enumerated.  
If $L1=L2=K$, the algorithm is called \emph{K-OPT} algorithm. 

{\bf Adopting Adequate Sub-graphs to Simplify Problem Space}
By using the adequate subgraphs \cite{weixu1,weixu3}, we can prove that they are still applicable for decomposing the graph in the \emph{DCJ-Indel-Exemplar(Matching)} median problem.
\begin{lemma}
As long as the irregular vertices do not involve, regular subgraphs are applicable to decompose \emph{MBG}.
\end{lemma}

\begin{proof}
If there are $d$ number of vertices that contain duplicated edges in \emph{MBG}, 
we can disambiguate the \emph{MBG} by generating different subgraphs that contain only one of the duplicate edge. 
We call these subgraphs disambiguate \emph{MBG}, (\emph{d-MBG}),
and there are $O(\prod_{i<d}deg(i))$ number of \emph{d-MBG}s. 
If a regular adequate subgraph exists in the \emph{MBG}, it must also exists in every \emph{d-MBG}.
Based on the \emph{0-matching} solution, 
we can transform every \emph{d-MBG} into completed \emph{d-MBG} (\emph{cd-MBG}) by constructing the optimal completion 
\cite{phillip1} between \emph{0-matching} and all the other three types of edges. 
After this step, the adequate subgraphs in every \emph{d-MBG} still exist in every \emph{cd-MBG}, 
thus we can use these adequate subgraphs to decompose \emph{cd-MBG} for each median problem without losing accuracy. $\Box$
\end{proof}

\begin{algorithm}[t]
\KwIn{\emph{MBG} $G$, Search Level $L1$ and $L2$}
\KwOut{\emph{0-matching} of $G$}
Init search list $L$ of size $L1$\;
Init \emph{0-matching} of $G$\;
$currentLevel \gets 0$ \ and {$Improved \gets true$\;
\While{$Improved = true$}{
    $currentLevel \gets 0$ and $Improved \gets false$\;
    Insert $G$ into $L[0]$\;
    \While{$currentLevel < L2$}{
         $G' \gets$ pop from list $L[currentLevel]$\;
         \If{$G'$ improves the median score}{
               $G \gets G'$\; $Improved \gets true$ and break \;
          }
            \If{$currentLevel < L1$} {
               \For{$x \in \forall$ $0$-$matching$ pairs of $G$}{
                   $G' \gets$ perform \emph{DCJ} on $G'$ using $x$\;
                   \lIf{$num\_pair(x) > \delta$}{
                        Insert $G'$ into $L[currentLevel+1]$
                    }
                    }
            }
            \Else{
                   $G' \gets$  perform \emph{DCJ} on $G'$ using $x=\underset{x} {\mathrm{argmax}} ~num\_pair(x)$ \;
                   \lIf{$num\_pair(x) > \delta$}{
                        Insert $G'$ into $L[currentLevel+1]$
                    }
            }
            $currentLevel \gets currentLevel +1$ \;
        }
    }
}
\Return{$0$-$matching$ of $G$}\;
\caption{{\sc DCJIndelExem(Matc)Median}}
\label{algo:median}
\end{algorithm}

{\bf Search Space Reduction Methods}
The performance bottleneck with the median computation is in the exhaustive search step, 
because for each search level we need to consider $O(|E|^2)$ possible number of edge pairs, which is $O(|E|^{2L1})$ in total.
Unlike the well-studied traveling salesman problem (\emph{TSP}) where it is cheap to find the best neighbor, 
here we need to compute the \emph{DCJ-Indel-Exemplar(Matching)} problem,\emph{NP}-hard distance, which makes this step extremely expensive to conclude.
Noticing that if we search neighbors on edges that are on the same \emph{0-i} color altered connected component (\emph{0-i-comp}),
the \emph{DCJ-Indel-Exemplar(Matching)} distance for genome 0 and genome $i$ is more likely to reduce \cite{YinTSB13}, 
thus we can sort each edge pair by how many \emph{0-i-comp} they share.
Suppose the number of \emph{0-i-comp} that an edge pair $x$ share is $num\_pair(x)$, 
when the algorithm is in the exhaustive search step ($currentLevel<L1$),
we set a threshold $\delta$ and select the edge pairs that satisfy $num\_pair(x)>\delta$ to add into the search list.
When it comes to the recursive deepening step, we select the edge pair that satisfy
$\underset{x} {\mathrm{argmax}} ~num\_pair(x)$ to add into the search list.
This strategy has two merits: 1) some of the non-promising neighbor solution is eliminated to reduce the search space; 
2) the expensive evaluation step which make a function call to \emph{DCJ-Indel-Exemplar(Matching)} distance 
is postponed to the time when a solution is retrieved from the search list.

The \emph{LK} based median computation algorithm is as Algorithm~\ref{algo:median} shows.

\section{Experimental Results}
We implement our code with python and C++: the python code realized the optimization methods 
while the C++ code is implemented on a parallel branch-and-bound framework \emph{OPTKit}.
We conduct extensive experiments to evaluate the accuracy and speed of our distance and median algorithms using both simulated and real biological data.
Experimental tests ran on a machine with linux operating system configured with 16
Gb of memory and an Intel(R) Xeon(R) CPU E5530 16 core processor, each core has 2.4GHz of speed.
All of the experiments ran with a single thread.
We choose to use g++-4.8.1 as our compiler.

\begin{figure*}[ht!]
	\begin{center}
   		\subfigure[Time result for simulated data.]{
	 	\label{med1}
	   	\includegraphics[width=0.52\textwidth]{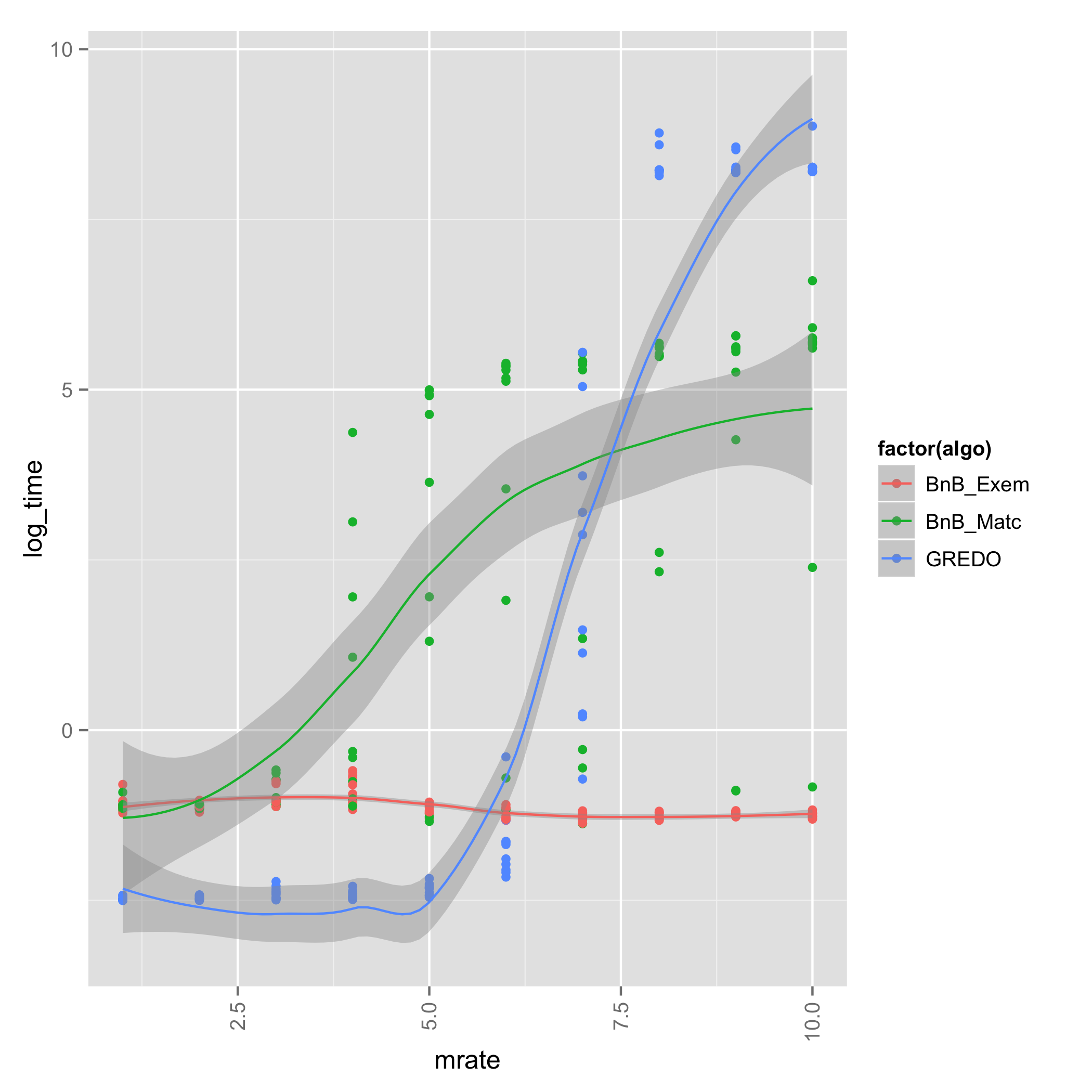}
   		}%
		\subfigure[Distance result for simulated data.]{
		\label{med2}
		\includegraphics[width=0.52\textwidth]{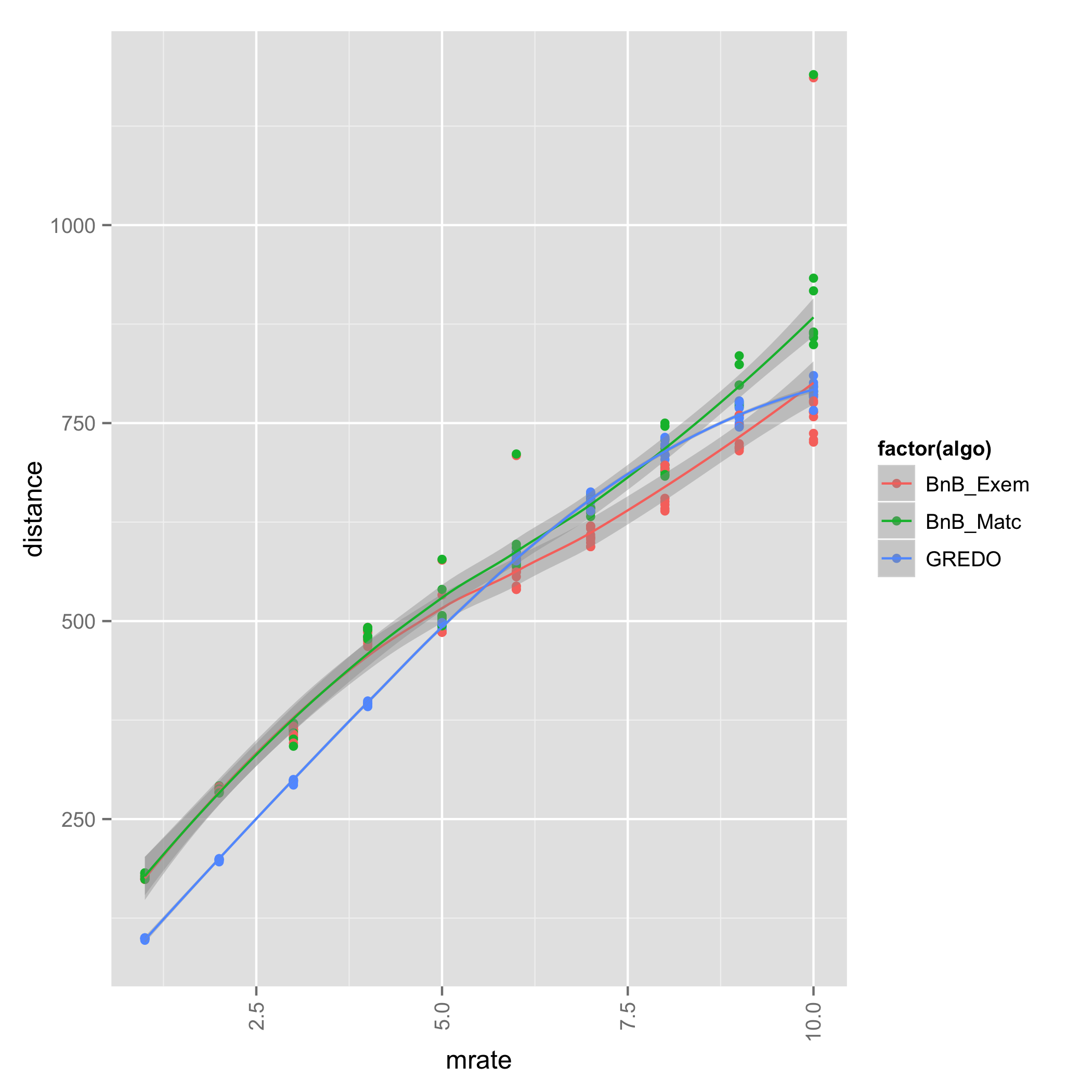}
		}
	\end{center}
	\caption{
		Experimental results for distance computation using simulated data.
	}
\label{disres}
\end{figure*}

\begin{table}[h]
\begin{center}
\begin{tabular}{|c|c|c|c|c|c|c|}
\hline
                  & \multicolumn{3}{c|}{Distance Results} & \multicolumn{3}{c|}{Time Results} \\ \hline
                  Data              & GREDO     & Exem        & Matc        & GREDO     & Exem      & Matc  \\ \hline
                  brownrat\_chicken & 1678      & 24546       & 24704       & 3604.28   & 172.73    & 7.45      \\ \hline
                  brownrat\_gorilla & 1274      & 17922       & 17966       & 5707.13   & 12.64     & 12.10     \\ \hline
                  brownrat\_human   & 1083      & 17858       & 17900       & 3725.76   & 12.14     & 12.19     \\ \hline
                  brownrat\_mouse   & 790       & 15433       & 15445       & 3725.66   & 14.51     & 15.06     \\ \hline
                  chicken\_gorilla  & 1491      & 16379       & 16421       & 3725.62   & 7.54      & 7.57      \\ \hline
                  chicken\_human    & 1521      & 16231       & 16276       & 3725.65   & 7.74      & 7.47      \\ \hline
                  chicken\_mouse    & 1528      & 15712       & 15745       & 3726.03   & 9.82      & 8.16      \\ \hline
                  gorilla\_human    & 486       & 17798       & 17798       & 3607.63   & 13.94     & 13.81     \\ \hline
                  gorilla\_mouse    & 860       & 18914       & 18935       & 4816.31   & 12.60     & 12.13     \\ \hline
                  human\_mouse      & 749       & 18126       & 18144       & 94.64     & 12.45     & 12.61     \\ \hline
\end{tabular}
\end{center}
\vspace{3 mm}
    \caption{
	Experimental results for disntance computation with real data set.
     }
\label{realtable}
\end{table}

\begin{figure*}[ht!]
    \begin{center}
        \subfigure[$\gamma=\phi=0\%$ and $\theta$ varies from $10\%$ to $100\%$.]{
            \label{med1}
            \includegraphics[width=0.55\textwidth]{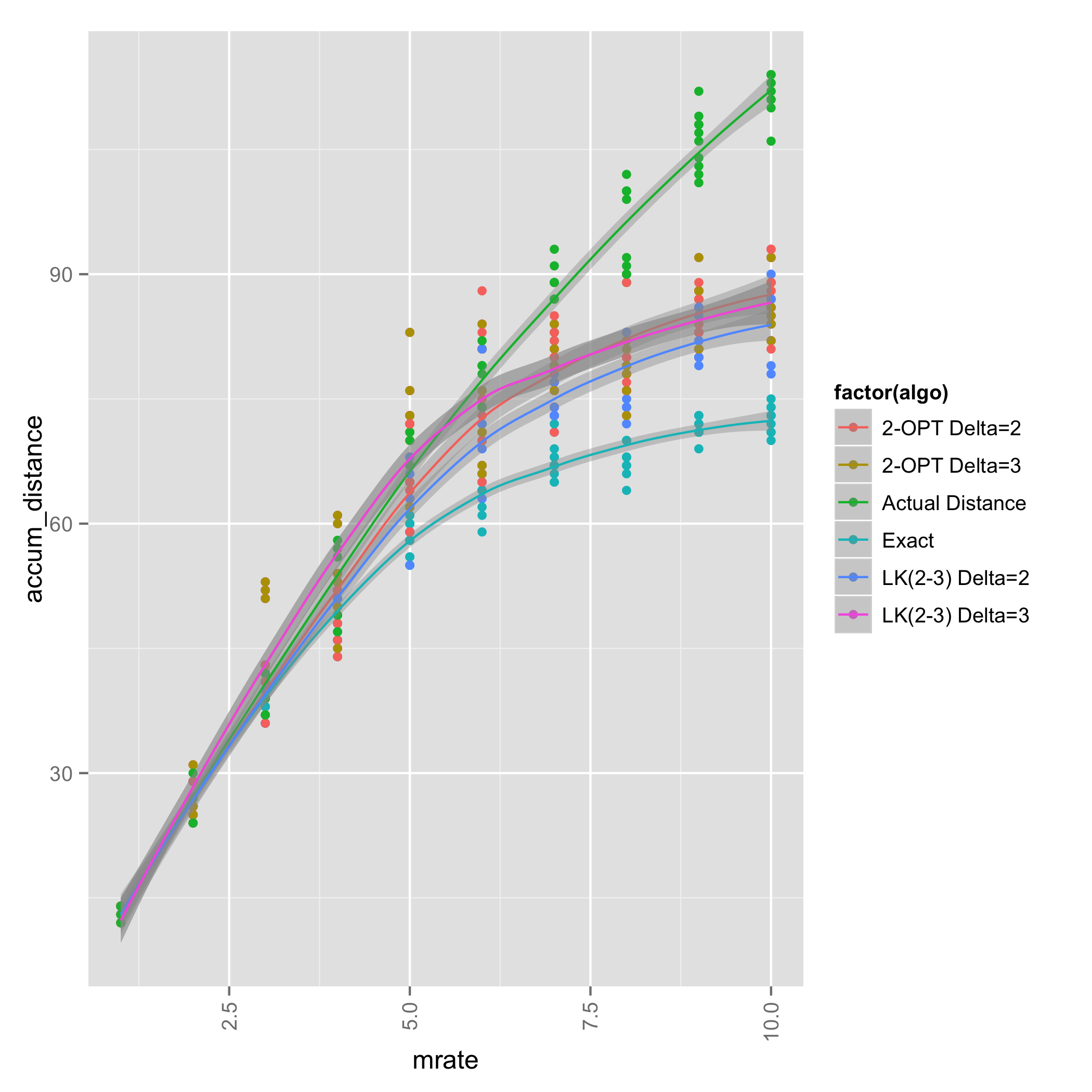}
        }%
        \subfigure[$\gamma=\phi=5\%$ and $\theta$ varies from $10\%$ to $60\%$.]{
           \label{med2}
           \includegraphics[width=0.55\textwidth]{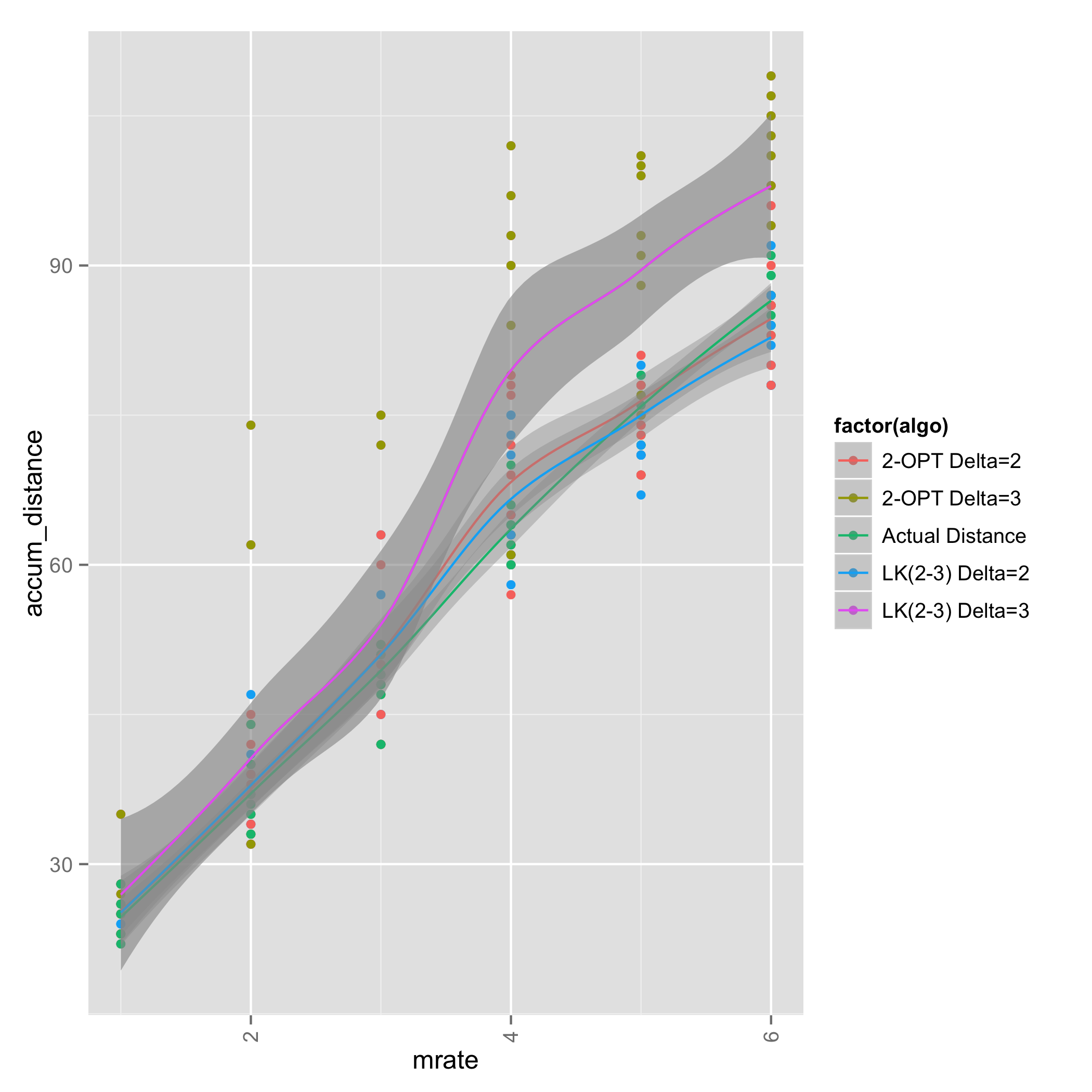}
        }
    \end{center}
    \caption{
	Experimental results for median computation applying \emph{DCJ-Indel-Exemplar} distance.
     }
\label{medexem}
\end{figure*}

\begin{figure*}[ht!]
    \begin{center}
        \subfigure[$\gamma=\phi=5\%$ and $\theta$ varies from $10\%$ to $100\%$.]{
            \label{med3}
            \includegraphics[width=0.55\textwidth]{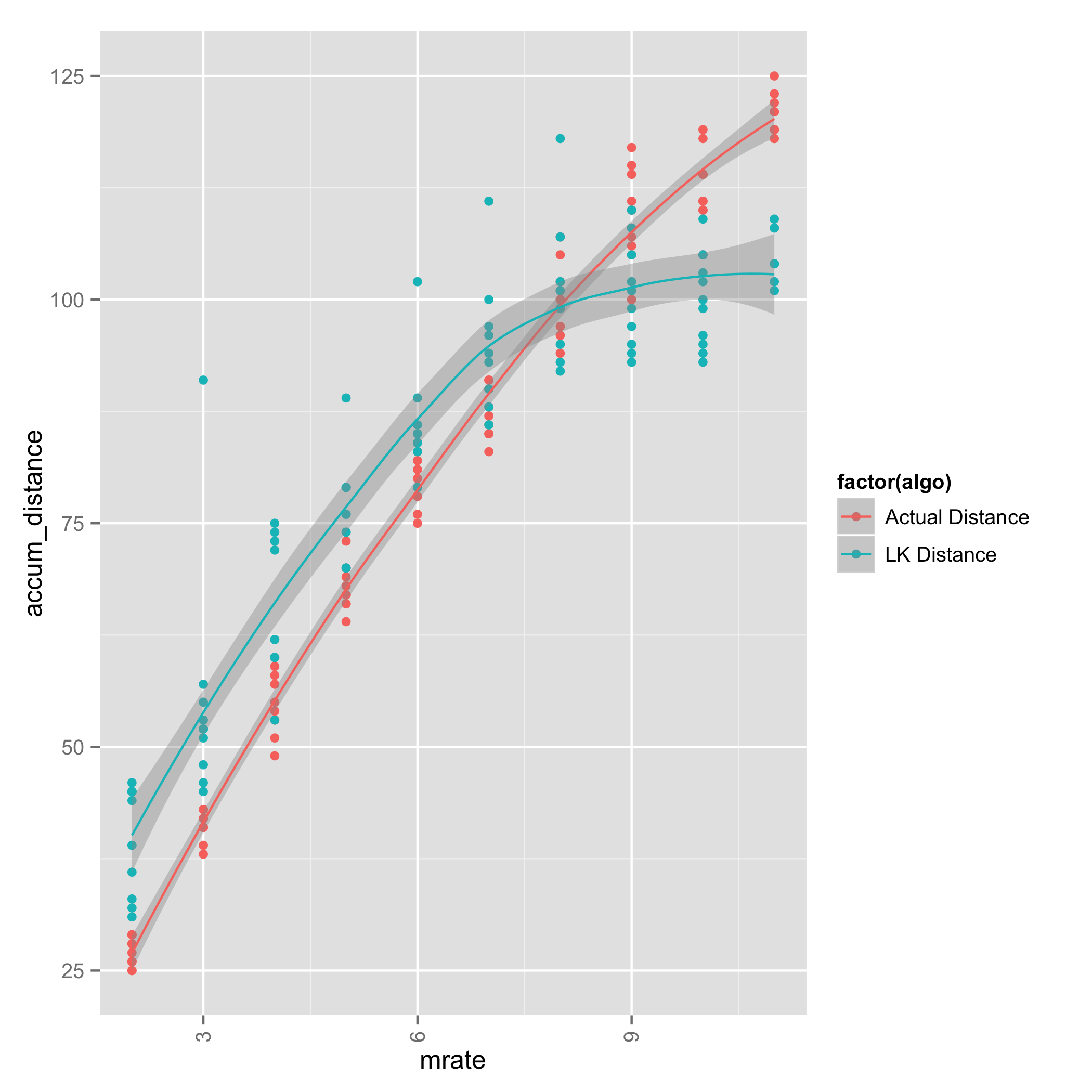}
        }%
        \subfigure[$\gamma=\phi=10\%$ and $\theta$ varies from $10\%$ to $100\%$.]{
           \label{med4}
           \includegraphics[width=0.55\textwidth]{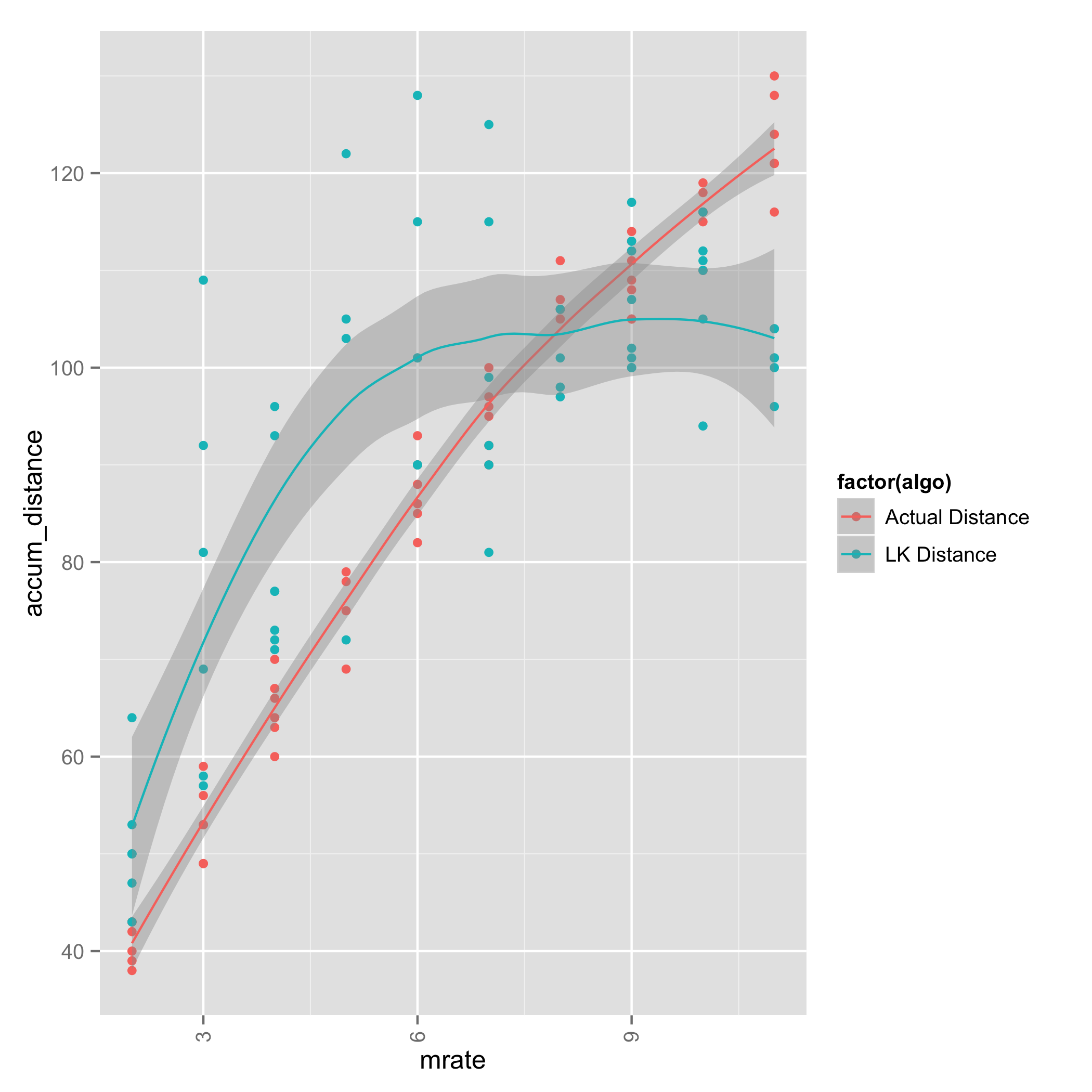}
        }
    \end{center}
    \caption{
	Experimental results for median computation applying \emph{DCJ-Indel-Matching} distance.
     }
\label{medexem}
\end{figure*}

\subsection{Distance Computation}
To the best of our knowlege, there is no software package that can handle both duplications and \emph{Indels}.
We compare our \emph{DCJ-Indel-Exemplar (Matching)} distances with \emph{GREDO} \cite{shao13}, a software package based on linear programming that can handle duplications.

{\bf Simulated Data} 
The simulated data sets are generated with genomes containing 1000 genes. 
The \emph{Indels} rate is set ($\gamma$) as $5\%$,  
inline with the duplication rate ($\phi$) as $10\%$. 
Considering \emph{GREDO} can not process \emph{Indel} data, all \emph{Indels} for \emph{GREDO} are removed. 
We compare the change of distance estimation with the variation of mutation rate ($\theta$, which grows from $10\%$ to $100\%$.
The experimental results for simulated data are displayed in  Figure~\ref{disres}.

\begin{enumerate}
\item For computational time, since the results of time spans over a range of thousands of seconds, we display the time with log scale to construe results clearly. 
When the mutation rate is less than $50\%$, all three methods perform similarly, with the fact that \emph{GREDO} is faster than both of our branch-and-bound methods. 
However, \emph{GREDO} slows down dramatically when the mutation rate is increased, 
while our branch-and-bound based method takes less increased time to finish.  
\item For computational accuracy, we show the distance results corrected by \emph{EDE} approach which is one of the best true distance estimator. 
As for simulated data, we can see that when the mutation rate is small (< $50\%$) \emph{GREDO} under estimate the distance as opposed to our two branch-and-bound methods;
but it will over estimate the distance with the growth of mutation rate. 
\end{enumerate}

{\bf Real data} 
We prepare the real data sets using genomes downloaded from Ensenble and processed them following the instructions in \cite{shao13}.
The real data set contains 5 species: brown-rat, chicken, human, mouse and gorilla.
For \emph{DCJ-Indel-Exemplar (Matching)} distance, we only convert the Ensenmble format to adapt the data to our program.
Meanwhile, just as the simulated data, all \emph{Indels} in real data set for \emph{GREDO} are removed.
The results for real data are shown in Table~\ref{realtable}.

\begin{enumerate}
\item For computational time, the branch-and-bound method shows orders of magnitudes of speed up compared with \emph{GREDO}. 
We analyze the data, the reason can be construed as the existance of multiple connected comonent in \emph{BPG}.
So that our method can divide the graph into much smaller size, versus \emph{GREDO} which doesn't have this mechanism.  
\item For computational accuracy, the distance results of the real data gives us a taste of how frequently \emph{Indels} happend in the genome evolution.  
We can see orders of magnitude of difference between our distance results and \emph{GREDO}, 
which is mainly due to the large amount of \emph{Indels} in the real data set.
Note that we did not change the way \emph{GREDO} compute its distance as in paper \cite{shao13}, 
in the real distance computation, we should consider \emph{Indels} in alignment with duplications. 
\end{enumerate}

\subsection{Median Computation}
{\bf Median Computation}
We simulate the median data of three genomes using the same strategy as in the distance simulation.
In our experiments, each genome is ``evolved'' from a seed genome, 
which is identity, and they all evolve with the same evolution rate ($\theta$, $\gamma$ and $\phi$).
The sequence length in the median experiments are reduced to 50, due to performance issues.

{\bf \emph{DCJ-Indel-Exemplar} median}  
We analyze the result of using \emph{LK} algorithm  with $L1=2$ and $L2=3$, and the \emph{K-OPT} algorithm of $K=2$.
Search space reduction methods are used, with $\delta=2$ and $\delta=3$ respectively.

\begin{enumerate}
\item To begin with, we compare our result along with equal content data, since there are already benchmark programs to help us performing analysis.
We run the exact \emph{DCJ} median solver (we use the one in \cite{YinTSB13}) to compare our heuristic with the exact median results.
In Fig~\ref{med1}, it shows the accuracy of our heuristic versus the exact result.
It is shown that when $\theta \leq 60\%$, all results of the \emph{LK} and \emph{K-OPT} methods are quite close to the exact solver.
For parameter of $\delta=2$, both \emph{LK} and \emph{K-OPT} methods can generate exactly the same results for most of the cases.
\item As for the median results for unequal contents, we set both $\gamma$ and $\phi$ to $5\%$ and increase the mutation (inversion) rate $\theta$ from $10\%$ to $60\%$.
We compare our results with the accumulated distance of the three genomes to their simulation seed. 
Although it can not show the accuracy of our method (since we do not have an exact solver),
it can be used as an indicator of how close that our method is to the real evolution.
Fig~\ref{med2} shows that when $\delta=3$, both the \emph{LK} and \emph{K-OPT} algorithms get results quite close to the real evolutionary distance.
\end{enumerate}

{\bf \emph{DCJ-Indel-Matching} median}  
Since \emph{DCJ-Indel-Exemplar} median has already given us the result of how \emph{LK} performs against exact solver, 
and how different parameters of \emph{LK} performs. 
With these things in mind, we choose to use \emph{LK} with $L1=2$ and $L2=3$ having $\delta=2$ as the configuration for our \emph{DCJ-Indel-Matching} median solver.
We use the same data as in the previous experiments, 
and the experimental results are shown in Figure~\ref{med3} and Figure~\ref{med4}.
We can see that in general, the new implementation is quite close to the real result when $\gamma=5\%$ 
and $\phi=5\%$ and slightly worse than real result when $\gamma=10\%$ and $\phi=10\%$.

\section{Conclusion}
In this paper, we proposed a new way to compute the distance and median between genomes with unequal contents (with \emph{Indels} and duplications). 
Our distance method can handle Indels which is ubiquitous in the real data set, and is proved to be more efficient as opposed to \emph{GREDO}. 
We designed a Lin-Kernighan based method to compute median, which can get close to optimal results in alignment with the exact median solver, 
and our methods can handle duplications and \emph{Indels} as well.

\section{Acknowledgements}
This Research was sponsored in part by the NSF OCI-0904461 (Bader), OCI-0904179, IIS-1161586 (Tang) and OCI-
0904166 (Schaeffer).

\bibliography{llncs}
\end{document}